\documentclass{article}
\usepackage[T1]{fontenc} 
\usepackage[utf8]{inputenc} 
\usepackage{ismir,amsmath,cite,url}
\usepackage{graphicx}
\usepackage{color}
\def\lbd{}	

\usepackage{booktabs}

\title{Beyond chord vocabularies: Exploiting pitch-relationships in a chord estimation metric}

\oneauthor
 {Johanna Devaney}
 {Brooklyn College and the Graduate Center, CUNY \\ {\tt johanna.devaney@brooklyn.cuny.edu}}

\def\authorname{J. Devaney}

\begin{document}

\maketitle
\begin{abstract}
Chord estimation metrics treat chord labels as independent of one another. This fails to represent the pitch relationships between the chords in a meaningful way, resulting in evaluations that must make compromises with complex chord vocabularies and that often require time-consuming qualitative analyses to determine details about how a chord estimation algorithm performs. This paper presents an accuracy metric for chord estimation that compares the pitch content of the estimated chords against the ground truth that captures both the correct notes that are estimated and additional notes that are inserted into the estimate. This is not a stand-alone evaluation protocol but rather a metric that can be integrated as a weighting into existing evaluation approaches.
\end{abstract}
\section{Introduction}\label{sec:introduction}

Chord estimation has a long history at ISMIR \cite{Pauwels19}, yet even current approaches still have not exceeded 90\% accuracy on simple chord label prediction tasks. Part of the challenge of chord label classification is the large number of label permutations \cite{mcfee17} and annotator subjectivity for rare chords, as well as idiosyncratic annotation styles \cite{koops17} and disagreements between expert annotators \cite{humphrey15,condit18,koops19}. Another part is that the evaluation metrics typically employed treat chord labels as independent, rather than as collections, or sets, of pitches. Thus incorrect chord labels are treated the same in these metrics, regardless of whether or not they have any common pitch content with the ground truth. This paper presents an accuracy metric that can be integrated as a weighting mechanism with existing chord estimation evaluation approaches. The benefit of this approach is that it captures pitch relationships between the predicted chords and provides a more nuanced evaluation than treating the chord labels as independent from one another. 

\section{Background}
Chord-labels are a common harmonic representation in both the symbolic and audio domains. The use of chord labels in symbolic music comes out of a long tradition of Roman numeral-focused pedagogy, particularly in North America (e.g., \cite{riemann95,piston1948}). Models developed in music theory and cognition, such as Krumhansl’s \cite{krumhansl90} and Lerdahl’s \cite{lerdahl01} work, have informed the development of computational distance metrics for chords (e.g., \cite{crawford05, de08, bello05}). In contrast, much of the harmonic analysis work in the field of music information retrieval has focused on chord recognition because of its simple mapping to a classification problem. Current chord estimation metrics, as exemplified by the ones currently in use MiREX (shown in Table \ref{tab:eval}), focus on the prediction of chord labels, with varying degrees of simplifications applied to account for chord vocabulary size \cite{pauwels13}. 

\begin{table}
 \begin{center}
 \begin{tabular}{l l}
  \hline
1. & Chord root note only\\
2. & Major and minor: N, maj, min\\
3. & Seventh chords: N, maj, min, maj7, min7, 7\\
4. & Major and minor with inversions: N, maj, min, \\
 & maj/3, min/b3, maj/5, min/5\\
5. & Seventh chords with inversions: N, maj, min, maj7, \\
 & min7, 7, maj/3, min/b3, maj7/3, min7/b3, 7/3, maj/5, \\
 & min/5, maj7/5, min7/5, 7/5, maj7/7, min7/b7, 7/b7\\
  \hline
 \end{tabular}
\end{center}
 \caption{List of the chord vocabulary classes, based on \cite{pauwels13}, that are used for evaluation in the current MIREX Audio Chord Estimation task.}
 \label{tab:eval}
\end{table}

One problem with the formulation of chord estimation as a simple classification task with independent labels is that overlapping pitch content between chords is ignored.  The chord labels themselves (e.g., C d- e- F G a b\^{o} or I ii iii IV V vi vii\^{o})
do not themselves provide information about the relationship between the chords. However, their pitch content can be extracted and compared in order to reveal that, for example, the chords d- (ii in C Major) and F (IV in C Major) are musically closer, due to their shared pitch content, than F (IV) and G (V), even though F and G are closer to one another in the label set. The broader issue of conceiving harmonic analysis in terms of labels rather than pitch content has previously been discussed in \cite{devaney15} and \cite{kaliakatsos15}.

\begin{table*}[h]
\begin{center}
\begin{tabular}{ccccccccccccc}
\toprule
\textbf{note name} & C & C\# & D & D\# & E & F & F\# & G & G\# & A & A\# & B\\
\textbf{pitch class} & 0 & 1 & 2 & 3 & 4 & 5 & 6 & 7 & 8 & 9 & 10 & 11\\
\midrule
C (I)       & o & - & - & - & o & - & - & o & - & - & - & - \\ 
d (ii)      & - & - & o & - & - & o & - & - & - & o & - & - \\ 
e (iii)     & - & - & - & - & o & - & - & o & - & - & - & o\\ 
F (IV)      & o & - & - & - & - & o & - & - & - & o & - & - \\ 
G(V)        & - & - & o & - & - & - & - & o & - & - & - & o\\ 
a (vi)      & o & - & - & - & o & - & - & - & - & o & - & - \\ 
b$^{o}$ (vii$^{o}$) & - & - & o & - & - & o & - & - & - & - & - & o\\
\bottomrule
\end{tabular}
\caption{Summary of the pitch content in the diatonic triads in the C Major scale. Note names are listed in the top row and the numbers in the second row represent the 12 pitch classes. In the lower part of the
table, a o indicates the presence of a pitch class in a triad. In this representation, it is clear that d (ii) and F (IV) are musically closer to each other than F (IV) and G (V) because they share more pitch classes in common, i.e., F and A.}
\label{table:chords}
\end{center}
\end{table*}

Recent work, such as \cite{carsault18}, has attempted to leverage the overlap in pitch content between varying types of chords. Although this has been limited to defining chord alphabets amongst different chord qualities or types (such as major, minor, or diminished) for a single chord rather than between different chords.

\section{Proposed accuracy metric}

In any of the evaluation classes shown in Table \ref{tab:eval}, if an F major chord (IV in C major) is misestimated as a d minor chord (ii), it would be considered equally incorrect as if it were misestimated as a G major chord (V). This fails to capture the fact there are two of the three notes in common between d minor and F major chords, while no notes are in common between the G major and F major chords, as shown in Table \ref{table:chords}. Chord labels are short-hand for pitch collections and treating them as independent labels in evaluation metrics makes it much harder to decode where an algorithm is succeeding and failing since all errors are weighted equally.

The following accuracy metric for chord estimation evaluation can be added as a weighting to existing evaluation approaches and is applicable both in the audio and symbolic domains.\footnote{An implementation of the metric is available at \url{https://github.com/jcdevaney/chordEstimationMetric}.} Let $C$ be the number of predicted notes $\hat{y}$ in the ground truth correctly identified $y$
\begin{equation}\label{relativity}
C = | y \cap \hat{y} |
\end{equation}
Let $I$ be the number of insertions (extra predicted notes) in the estimated chord that are not present in the ground truth.
\begin{equation}\label{relativity}
I = | \hat{y} \backslash y |
\end{equation}
Let $A$ be the accuracy measurement for each chord estimate, calculated from $C$ and $I$ scaled between 0 and 1. 
\begin{equation}\label{relativity}
A = \frac{C - I+|y|}{2|y|}
\end{equation}
Thus, $A$ provides a combined measurement of which notes are correctly predicted and whether any additional notes (insertions), in excess of the number of notes in the ground truth chord, were predicted.

Using the example from above of F chord misestimated as either a d- chord or a G chord, we can see how this accuracy measurement captures the differences in the pitch content between the two estimates (using the pitch class content in Table \ref{table:chords}). To calculate the accuracy for d-, $A_d$, we will compare the pitch classes of d-, \{2,5,9\}, to those of F, \{0,5,9\}.
\begin{equation}\label{relativity}
C_d = | \{0,5,9\} \cap \{2,5,9\} | \\
\end{equation}
\begin{equation}\label{relativity}
C_d = 2
\end{equation}
\begin{equation}\label{relativity}
I_d = |\{2,5,9\} \backslash \{0,5,9\} |
\end{equation}
\begin{equation}\label{relativity}
I_d = 1
\end{equation}
\begin{equation}\label{relativity}
A_d = \frac{2 - 1+|3|}{2|3|}
\end{equation}
\begin{equation}\label{relativity}
A_d = 0.66
\end{equation}

To calculate the accuracy for G, $A_G$, we will compare the pitch classes of G, \{2,7,11\}, to those of F, \{0,5,9\}.
\begin{equation}\label{relativity}
C_G = | \{0,5,9\} \cap \{2,7,11\} |
\end{equation}
\begin{equation}\label{relativity}
C_G = 0
\end{equation}
\begin{equation}\label{relativity}
I_g = |\{0,5,9\} \backslash \{2,7,11\}|
\end{equation}
\begin{equation}\label{relativity}
I_g = 3
\end{equation}
\begin{equation}\label{relativity}
A_g = \frac{0 - 3+|3|}{2|3|}
\end{equation}
\begin{equation}\label{relativity}
A_g = 0
\end{equation}

\section{Conclusions}
This proposed accuracy metric works directly on the pitch class information and can help to facilitate examinations of where chord estimation algorithms provide partially correct answers. Such an examination can lead to a more nuanced understanding of the algorithms and more efficient algorithm refinement. It can either be used as a weighting with existing evaluation approaches or further developed to replace the complex chord vocabularies currently in use.   


\newpage
\bibliography{ISMIRtemplate}

%
%
%
%
%

\end{document}